\documentclass[%
amsmath,amssymb,prl,superscriptaddress,twocolumn
]{revtex4-2}

\usepackage{graphicx}
\usepackage{dcolumn}
\usepackage{bm}
\usepackage{hyperref}
\usepackage{natbib}
\usepackage{float}
\usepackage{amsmath}
\usepackage{xcolor}

\newcommand{\red}[1] {{\color{black}#1}}
\newcommand{\redd}[1] {{\color{black}#1}}

\newcommand{\beginsupplement}{
    \onecolumngrid
    \setcounter{page}{1}
    \setcounter{table}{0}
    \renewcommand{\thetable}{S\arabic{table}}
    \setcounter{figure}{0}
    \renewcommand{\thefigure}{S\arabic{figure}}
    \setcounter{equation}{0}
    \renewcommand{\theequation}{S\arabic{equation}}
    \setcounter{secnumdepth}{2}
}

\begin{document}

\title{Microwave-assisted unidirectional superconductivity in Al-InAs nanowire-Al junctions under magnetic fields}

\author{Haitian Su}
\affiliation{Beijing Key Laboratory of Quantum Devices, Key Laboratory for the Physics and Chemistry of Nanodevices, and School of Electronics, Peking University, Beijing 100871, China}
\affiliation{Institute of Condensed Matter and Material Physics, School of Physics, Peking University, Beijing 100871, China}

\author{Ji-Yin Wang}
\affiliation{Beijing Academy of Quantum Information Sciences, Beijing 100193, China}

\author{Han Gao}
\affiliation{Beijing Key Laboratory of Quantum Devices, Key Laboratory for the Physics and Chemistry of Nanodevices, and School of Electronics, Peking University, Beijing 100871, China}

\author{Yi Luo}
\affiliation{Beijing Key Laboratory of Quantum Devices, Key Laboratory for the Physics and Chemistry of Nanodevices, and School of Electronics, Peking University, Beijing 100871, China}
\affiliation{Institute of Condensed Matter and Material Physics, School of Physics, Peking University, Beijing 100871, China}

\author{Shili Yan}
\affiliation{Beijing Academy of Quantum Information Sciences, Beijing 100193, China}

\author{Xingjun Wu}
\affiliation{Beijing Academy of Quantum Information Sciences, Beijing 100193, China}

\author{Guoan Li}
\affiliation{Beijing National Laboratory for Condensed Matter Physics, Institute of Physics, Chinese Academy of Sciences, Beijing 100190, China}
\affiliation{School of Physical Sciences, University of Chinese Academy of Sciences, Beijing 100049, China}

\author{Jie Shen}
\affiliation{Beijing National Laboratory for Condensed Matter Physics, Institute of Physics, Chinese Academy of Sciences, Beijing 100190, China}
\affiliation{Songshan Lake Materials Laboratory, Dongguan 523808, China}
\affiliation{Beijing Academy of Quantum Information Sciences, Beijing 100193, China}

\author{Li Lu}
\affiliation{Beijing National Laboratory for Condensed Matter Physics, Institute of Physics, Chinese Academy of Sciences, Beijing 100190, China}
\affiliation{School of Physical Sciences, University of Chinese Academy of Sciences, Beijing 100049, China}
\affiliation{Songshan Lake Materials Laboratory, Dongguan 523808, China}

\author{Dong Pan}
\email{pandong@semi.ac.cn}
\affiliation{State Key Laboratory of Superlattices and Microstructures, Institute of Semiconductors,Chinese Academy of Sciences, P.O. Box 912, Beijing 100083, China}

\author{Jianhua Zhao}
\affiliation{State Key Laboratory of Superlattices and Microstructures, Institute of Semiconductors,Chinese Academy of Sciences, P.O. Box 912, Beijing 100083, China}

\author{Po Zhang}
\email{zhangpo@baqis.ac.cn}
\affiliation{Beijing Academy of Quantum Information Sciences, Beijing 100193, China}

\author{H. Q. Xu}
\email{hqxu@pku.edu.cn}
\affiliation{Beijing Key Laboratory of Quantum Devices, Key Laboratory for the Physics and Chemistry of Nanodevices, and School of Electronics, Peking University, Beijing 100871, China}
\affiliation{Beijing Academy of Quantum Information Sciences, Beijing 100193, China}

\begin{abstract}
Under certain symmetry-breaking conditions, a superconducting system exhibits asymmetric critical currents, dubbed the ``superconducting diode effect". 
Recently, systems with the ideal superconducting diode efficiency or unidirectional superconductivity have received considerable interest. 
In this work, we report the study of Al-InAs nanowire-Al Josephson junctions under microwave irradiation and magnetic fields. 
We observe an enhancement of \red{superconducting diode effect} under microwave driving, featured by a horizontal offset of the zero-voltage step in the voltage-current characteristic that increases with microwave power. 
Devices reach the \red{unidirectional superconductivity} regime at sufficiently high driving amplitudes. 
The offset changes sign with the reversal of the magnetic field direction. Meanwhile, the offset magnitude exhibits a roughly linear response to the microwave power in dBm when both the power and the magnetic field are large.
The signatures observed are reminiscent of a recent theoretical proposal using the resistively shunted junction (RSJ) model. However, the experimental results are not fully explained by the RSJ model, indicating a new mechanism for \red{unidirectional superconductivity} that is possibly related to non-equilibrium dynamics \red{or dissipation} in periodically driven superconducting systems. 
\end{abstract}

\maketitle

{\em Background.}---
Directional behaviors are ubiquitous in nature. Examples are ratchets and diodes used in daily life. A quantum ratchet phenomenon gaining widespread interest recently is the superconducting diode effect (SDE)~\cite{yanson1965experimental, goldman1967meissner, ando2020observation, yuan2022supercurrent, daido2022intrinsic, ilic2022theory,he2022phenomenological, fominov2022asymmetric, souto2022josephson, nadeem2023superconducting,hou2023ubiquitous,Zhang2020nc,Pal2022np,Jeon2022nm,Satchell2023jap,Ryohei2017sa,Levichev2023prb,Ustavschikov2022JETP}. The SDE describes asymmetric critical currents in superconducting systems. It is related to symmetry-breaking physics and has potential application in building low-dissipation logical devices. This effect is also used to probe exotic systems such as $\phi_0$ junctions~\cite{zhang2022evidence,Szombati2016} and topological edge states~\cite{li2023interfering}.

The SDE has been studied in superconductors or Josephson junctions made from a variety of materials~\cite{ando2020observation, lin2022zero, narita2022field, zhang2022evidence, mazur2022gate, turini2022josephson, wu2022field, baumgartner2022effect, baumgartner2022supercurrent, lotfizadeh2023superconducting, diez2023symmetry}. Origins of the SDE range from extrinsic factors, such as self-inductance or trapped vortexes~\cite{yanson1965experimental, goldman1967meissner, jiang1994asymmetric, ichikawa1994asymmetric, Suri2022apl, Golod2022nc, Gutfreund2023nc}, offset current from an external current source~\cite{chiles2023nonreciprocal}, and nonequilibrium driving~\cite{daido2023unidirectional}, to intrinsic ones, like the spin-orbit coupling and valley polarization~\cite{yuan2022supercurrent, daido2022intrinsic, hu2023josephson}. 
Superconducting quantum interference devices (SQUIDs) with high transparency or considerable loop inductance also exhibit the SDE with flux-tunable diode efficiencies~\cite{mayer2020gate, smash_second_harmonic, paolucci2023gate, ciaccia2023gate, Valentini2024}. The ideal SDE, or unidirectional superconductivity (USC), refers to the situation where the critical current ($I_c$) vanishes or becomes negative in the ``hard" direction. The two effects are usually observed together since the ideal SDE is a special kind of USC. 
The USC has been observed in the ``triode" structure consisting of three \red{Josephson junction}s~\cite{chiles2023nonreciprocal}, twisted trilayer graphene devices~\cite{lin2022zero}, and microwave-irradiated Al/Ge quantum well-based SQUIDs~\cite{Valentini2024}. Origins of USC are rather different in these systems, expected because the SDE itself can be due to a variety of mechanisms. In triode devices, the voltage-current characteristic is offset by an external current source. 
The mechanism for USC in the twisted trilayer graphene system remains to be understood while a nonequilibrium model is inspired by the experiment~\cite{daido2023unidirectional}. In microwave-driven SQUIDs, the USC is explained by the resistively shunted junction (RSJ) model with an SDE that already exists without the microwave~\cite{fominov2022asymmetric, souto2022josephson, souto2023tuning, PhysRevResearch.6.023011}. The study of the SDE or USC in periodically driven systems is still preliminary. We focus on such a system in our experiments.
 
\begin{figure}[H]\centering
\includegraphics{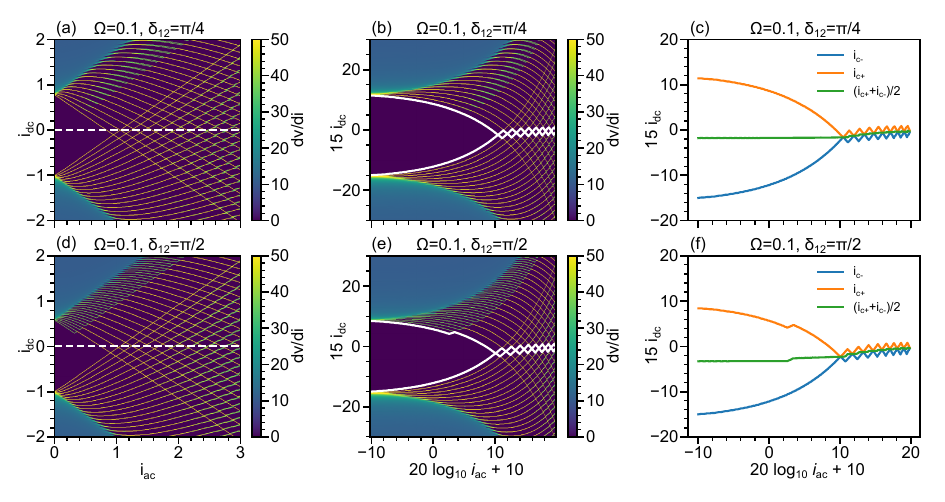}
\caption{\label{figS_analyse_sim} 
SDE simulated using the RSJ model. $\Omega = h f/ I_c R_n$ is the dimensionless frequency. $\delta_{12}$ is the relative phase offset between the first and second harmonic terms in the current-phase relation. When $i_{ac} = 0$, the diode efficiency roughly increases linearly as $\delta_{12}$ increases from 0 to $\pi/2$ (not shown). White curves in (b) and (e) are extracted $I_{c-}$ and $I_{c+}$, which are also shown in (c) and (f).
}
\end{figure}

For a \red{Josephson junction}, the SDE can be modelled with a toy current-phase relation: $I(\varphi) = I_1 \sin (\varphi+\varphi_0) + I_2 \sin (2 \varphi+2 \varphi_0 + \delta_{12})$, where $I$ is the supercurrent, $\varphi$ is the junction's phase difference, $\delta_{12}$ is the phase offset between two harmonic terms, $I_1$, $I_2$, and $\varphi_0$ are constant parameters~\cite{zhang2022evidence, fominov2022asymmetric, hu2023josephson, souto2022josephson, souto2023tuning}. 
A nonzero $\delta_{12}$ leads to
the SDE. This \red{current-phase relation} also works for SQUIDs. The current through a SQUID is described by a single phase difference because phase differences in separated junctions are interlocked by the magnetic flux threading the loop. Nonezero $\delta_{12}$ may be due to the interplay between spin-orbit coupling and magnetic fields~\cite{zhang2022evidence}, valley polarization~\cite{hu2023josephson}, or non-integer magnetic flux threading a SQUID~\cite{fominov2022asymmetric, souto2022josephson}. Substituting the \red{current-phase relation} into the RSJ model gives the USC~\cite{souto2022josephson,Valentini2024}. This can be understood as follows. For a system with SDE, the zero-voltage step is asymmetric about the origin. Two processes occur when the microwave amplitude increases: the size of the zero-voltage step shrinks and the center of the step moves toward the origin. The first process dominates at lower microwave amplitudes, leading to an increase in diode efficiency. The second process dominates at higher microwave amplitudes, suppressing the SDE (Fig.~\ref{figS_analyse_sim}).



\begin{figure}
\includegraphics{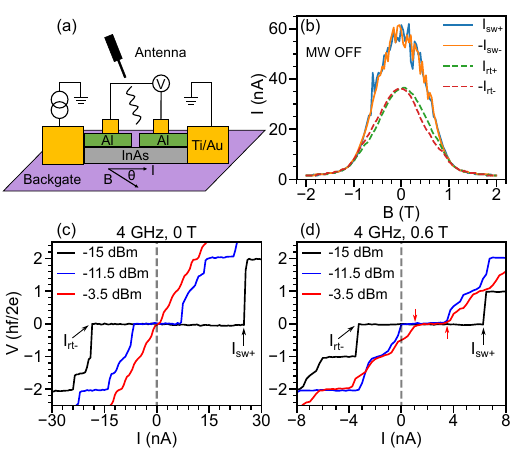}
\caption{\label{fig_overview}
Microwave-assisted superconducting diode effect in device A. (a) Schematic of the measurement. The in-plane magnetic field $B$ is applied via a solenoid, therefore $\theta$ is fixed. (b) Dependence of switching currents ($I_{sw+}$, $I_{sw-}$) and retrapping currents ($I_{rt+}$, $I_{rt-}$) on $B$. $B$ is scanned in the negative direction and the microwave is off. (c) and (d) Zero-field and finite-field voltage-current characteristics under microwave irradiation. The microwave frequency and $B$ are noted at the top of each panel. The current is scanned in the positive direction. Measured DC voltage are recalibrated by subtracting offsets. The backgate voltage is 0 V in (b) and (c), -0.5 V in (d).
}
\end{figure}

\begin{figure*}
\includegraphics{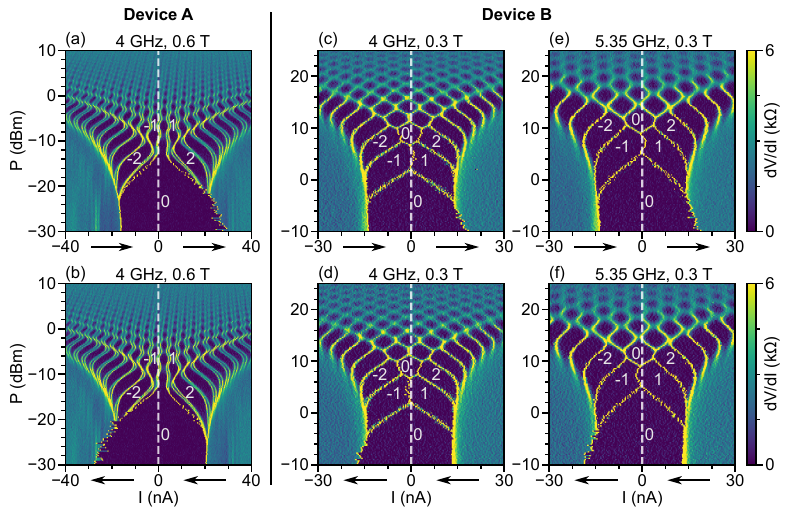}
\caption{\label{fig_freq_and_dev}
Differential resistance ($dV/dI$) maps in different current-scan directions for two devices. (a-b) Device A. The backgate voltage is $-0.5$ V. (c-f) Device B. The backgate voltage is $27$ V. $P$ and $I$ are the microwave power and the DC bias current, respectively. The microwave frequency and the magnet field are indicated at the top of each panel. Shapiro step indexes are labeled in white. Horizontal arrows indicate the scan direction of $I$.
}
\end{figure*}

{\em List of key results.}---
We study the SDE of Al-InAs nanowire \red{Josephson junction}s in the presence of microwave irradiation and magnetic fields.  Two devices (A and B) are studied and show similar results. The devices manifest weak SDE without microwave. An enhancement of the SDE is observed when junctions are subjected to microwave driving. The offset current of the zero-voltage step in the \red{voltage-current characteristic}, $I_{o\!f\!f}$, shifts away from the origin in microwave irradiation, leading to the USC regime at large microwave powers. At high magnetic fields, $I_{o\!f\!f}$ is reversed when the direction of the field flips, and roughly increases linearly with the microwave power in dBm at large power values. At lower magnetic fields, the response is less symmetric about the field. There is also weak zero-field non-symmetry between negative and positive retrapping currents, possibly due to accidentally trapped flux.

{\em Methods.}---
Devices are made from molecular beam epitaxy grown InAs nanowires covered by a layer of in-situ grown Al film (about 15 nm).
Details about the materials can be found in Ref.~\cite{pan2022situ}. Nanowires are transferred to the substrate randomly with a tissue. Junctions are formed by selectively wet etching of Al.
Measurements are performed in a dilution refrigerator with a base temperature about 15 mK~\cite{Yan2023}. 
The microwave is coupled to junctions via an antenna. The current-bias condition is assumed for the microwave signal because impedance of the microwave line and the air gap is much larger than the impedance of junctions~\cite{lehnert1999nonequilibrium}. Offsets of the order of 10~$\mu$V are subtracted from measured DC voltages. Origins of the offsets include the amplifier offset and thermal voltage drops in measurement lines.


{\em Figure~\ref{fig_overview} description.}---
Fig.~\ref{fig_overview} presents an example of the microwave-assisted USC in device A. The experimental setup is sketched in Fig.~\ref{fig_overview}(a). $\theta$ is $156^\circ$ ($68^\circ$) for device A (B)~\cite{sm}. The critical field of device A (B) is about 2~T (1~T).
Under microwave irradiation, \red{voltage-current characteristic}s show additional steps at finite voltages, which are Shapiro steps [Figs.~\ref{fig_overview}(c) and \ref{fig_overview}(d)]. In this work we focus on the zero-voltage step, or the zeroth Shapiro step in the context of AC driving.

\begin{figure}[H]\centering
\includegraphics{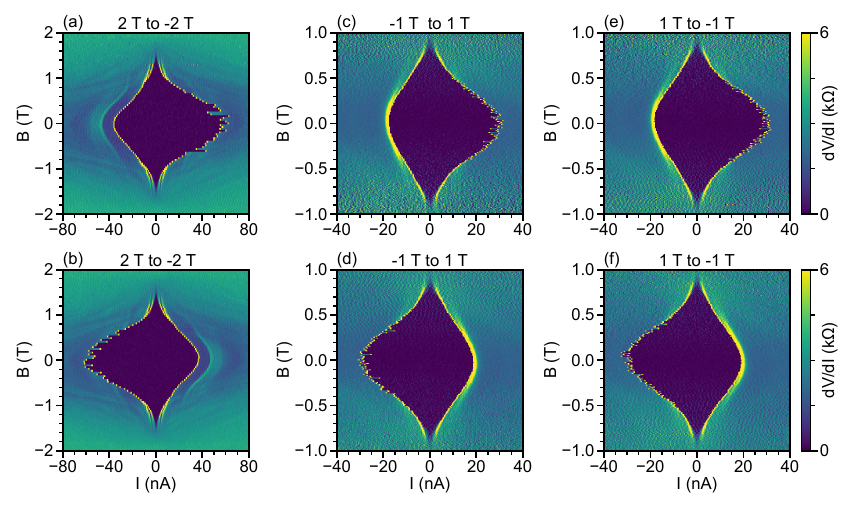}
\caption{\label{figS_field_2d_map} 
V-I characteristics in different magnetic field scan directions and current scan directions. (a-b) Device A at a backgate voltage of 0 V. (c-f) Device B at a backgate voltage of 27 V. Current is scanned in the positive direction in (a), (c), and (e), and in the negative direction in (b), (d), and (f). The magnetic field scan direction is indicated at the top of each panel. While there is a strong current hysteresis at low magnetic fields, no obvious magnetic hysteresis is observed.
}
\end{figure}

\begin{figure}[H]\centering
\includegraphics{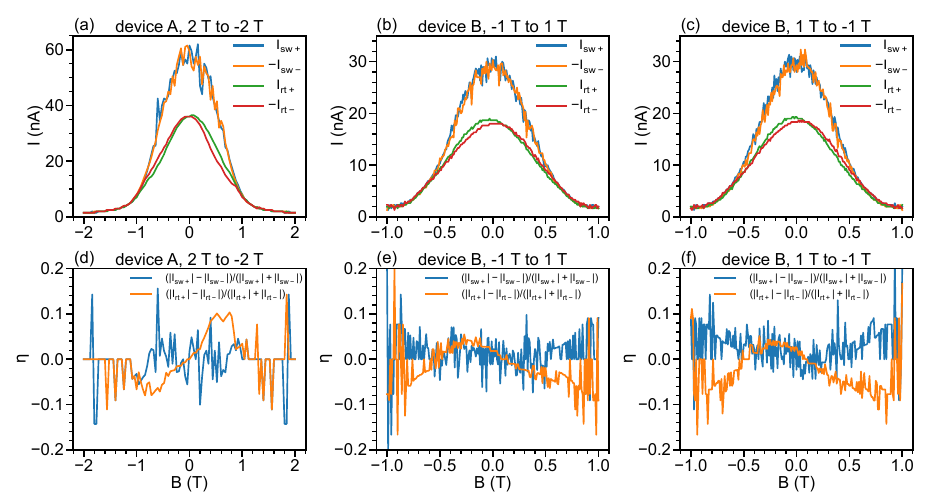}
\caption{\label{figS_dev23_diode} 
Extracted current parameters and diode efficiency from data in Fig.~\ref{figS_field_2d_map}. (a-c) Switching ($I_{sw+}$, $I_{sw-}$) and retrapping ($I_{rt+}$, $I_{rt-}$) currents as functions of $B$. (d-f) Diode efficiencies calculated using ($|I_{sw+}|-|I_{sw-}|)/(|I_{sw+}|+|I_{sw-}|$) and ($|I_{rt+}|-|I_{rt-}|)/(|I_{rt+}|+|I_{rt-}|$). Device name and magnetic field scan direction are indicated at the top of each panel.
}
\end{figure}

\begin{figure}[H]\centering
\includegraphics{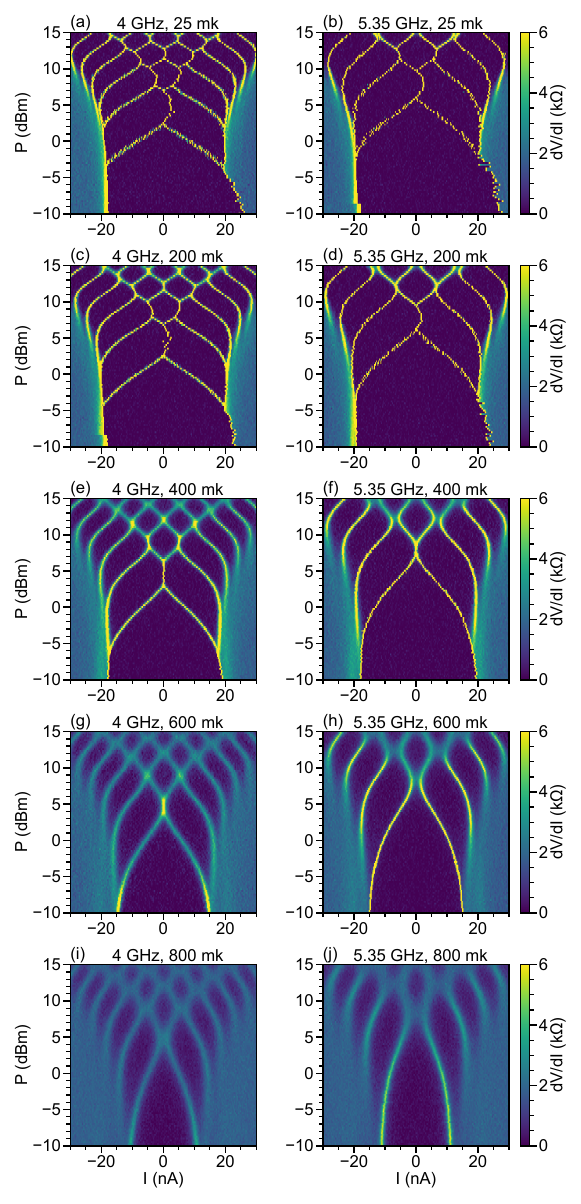}
\caption{\label{figS_DeviceB_ChangeT} 
\redd{Temperature dependence of Shapiro steps in Device B. The microwave frequency and temperature are indicated at the top of each panel. Current is sweeped in the positive direction. The switching current (positive branch) shows a stochastic behavior at low temperatures and low power values (see, e.g., panel (a)). This stochastic behavior is smeared out when the temperature or the microwave power increases. The difference in retrapping (negative branch) and switching (positive branch) currents, which indicates a hysteresis, is also weakened by raising the temperature or the microwave power.
The back gate voltage is 27 V.  The magnetic field is zero.}
}
\end{figure}

When the microwave is off, device A shows a weak SDE below 1 T [Fig.~\ref{fig_overview}(b)]. $I_{sw-}$ and $I_{sw+}$ ($I_{rt-}$  and $I_{rt+}$) are the negative and positive switching (retrapping) currents, extracted from \red{voltage-current characteristic}s scanned in the negative and positive (positive and negative) directions. For a single \red{voltage-current characteristic} scanned in one direction, two of the four parameters are extracted as indicated in Fig.~\ref{fig_overview}(c). 
The fluctuation in $I_{sw-}$ and $I_{sw+}$ at low fields is \redd{a stochastic behavior} due to premature transition to the normal state caused by \redd{electrical or flux} noise~\cite{10.1063/1.3543736, lotfizadeh2023superconducting,Trahms2023nature, Steiner2023PRL}. This phenomenon is pronounced when the critical current is large enough 
\redd{so the heating effect tends to trap the device in the normal state. The stochastic behavior is smeared out if the temperature increases (Fig.~\ref{figS_DeviceB_ChangeT}) or the critical current decreases (Fig.~\ref{figS_field_2d_map}), both of which reduces the electron temperature difference between the normal and superconducting states}. The fluctuation obscures the difference between $I_{sw-}$ and $I_{sw+}$ (if there is any). We \red{refer} the weak SDE \red{to} the difference between $I_{rt-}$ and $I_{rt+}$~\redd{\cite{lotfizadeh2023superconducting,Trahms2023nature, Steiner2023PRL}}. \red{While most SDE experiments focus on switching currents, retrapping currents can also have diode-like behavior although the origins may be different ~\cite{Trahms2023nature, Steiner2023PRL}}. There is also a slight asymmetry between $I_{rt-}$ and $I_{rt+}$ at $B = 0$. The zero-field asymmetry is more pronounced in device B (Fig.~\ref{figS_dev23_diode}), which may be due to accidentally trapped fluxes.

The difference between switching and retrapping currents indicates hysteresis. The hysteresis is also due to the heating effect as decreasing the critical current with the magnetic field or \red{increasing the temperature (Figs.~\ref{figS_field_2d_map} and \ref{figS_DeviceB_ChangeT})} reduces the hysteresis, \redd{ruling out the capacitance-effect explanation}~\cite{courtois2008origin,de2016interplay,shelly2020existence,dartiailh2021missing}. \redd{In the self-heating scenario, electrons in the normal and superconducting states have different temperatures, resulting in a bistable system with hysteresis. Increasing the temperature reduces the electron temperature difference in the two states, suppressing the hysteresis}. Under microwave irradiation and zero magnetic field [Fig.~\ref{fig_overview}(c)], the hysteresis is visible at -15 dBm, becoming negligible at -11.5 dBm and -3.5 dBm. If the field is set to 0.6 T, the device enters the ideal SDE regime and the USC regime at -11.5 dBm and -3.5 dBm, respectively [Fig.~\ref{fig_overview}(d)]. We note that in Fig.~\ref{fig_overview}(d) the current is scanned in the positive direction only. To obtain rigorous conclusion about the SDE, it is necessary to look at results from both scan directions like those in Fig.~\ref{fig_freq_and_dev}.

\begin{figure*}
\includegraphics{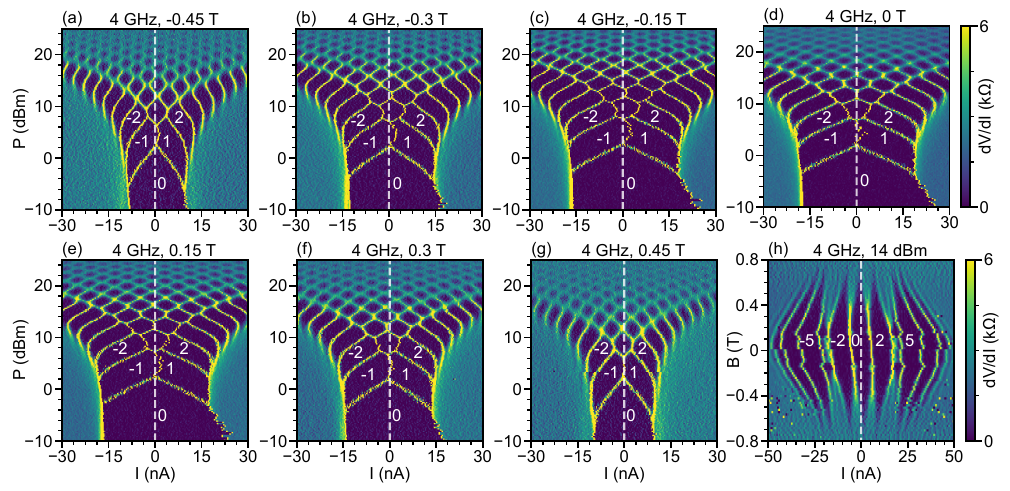}
\caption{\label{fig_magnet}
Field dependence of the microwave-assisted SDE in device B. (a-g) Differential resistance $dV/dI$ as a function of microwave power $P$ and DC current $I$. The microwave frequency and the magnetic field $B$ are indicated at the top of each panel. The back gate voltage is 27 V. (h)  $dV/dI$ as a function of $B$ and $I$. $B$ is scanned in the negative direction. Microwave power is 14 dBm. The back gate voltage is 30 V. Current is scanned in the positive direction in all panels. Shapiro step indexes are indicated in white.
}
\end{figure*}

\begin{figure}
\includegraphics{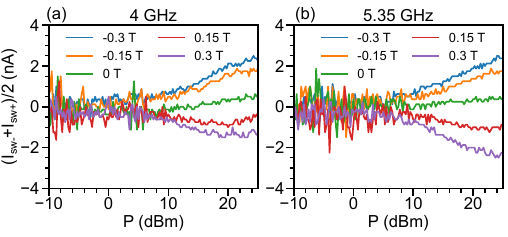}
\caption{\label{fig_analyse}
Offset current, $I_{o\!f\!f} = (I_{sw-} + I_{sw+})/2$, as a function of the microwave power in device B. $I_{sw-}$ and $I_{sw+}$ are extracted from datasets similar to those in Fig.~\ref{fig_magnet} and in both current-scan directions. (a) At 4 GHz. (b) At 5.35 GHz.
The backgate voltage is 27 V.
}
\end{figure}


{\em Figure~\ref{fig_freq_and_dev} description.}---
Detailed $dV/dI$ maps as a function of the microwave power $P$ and the DC current $I$ are shown in Fig.~\ref{fig_freq_and_dev}. In both devices A and B, the zero-voltage step (labeled as ``0") shows hysteresis at lower powers, i.e., below -20 dBm (-5 dBm) for device A (B). The current-scan direction is indicated by double arrows. The strong hysteretic regime is accompanied by fluctuations in $I_{sw}$. As discussed earlier, we attribute both phenomena to the heating effect. In the regime where $I_{sw}$ is significantly reduced by the microwave power, the hysteresis and fluctuation are much weaker. In device B, there is also weak hysteresis on the boundary between Shapiro steps -1 and 1 or 0 and $\pm 2$. The weak hysteresis can be explained by considering a shunted capacitor~\cite{larson2020zero}.

$dV/dI$ maps in Figs.~\ref{fig_freq_and_dev}(a) and \ref{fig_freq_and_dev}(b) show similar patterns above -15 dBm (where the oscillation of the zeroth Shapiro step about the power reaches its first node): the zeroth step shifts away from the origin in the same direction. At -5 dBm and higher power values, the whole zeroth step falls on the right side of $I = 0$ (vertical dashed lines) regardless of the current-scan direction. Similar trends are observed in device B at different microwave frequencies [Figs.~\ref{fig_freq_and_dev}(c)-\ref{fig_freq_and_dev}(f)]. The offset of the zeroth step in device B is opposite to that in device A due to a different $\theta$.

We define the offset current $I_{o\!f\!f} = (I_{sw-} + I_{sw+})/2$, which is the center of the zeroth step when the hysteresis is negligible. $I_{o\!f\!f}$ roughly increases linearly with the power above the zeroth step's first oscillation node. This is contrary to the RSJ scenario that $I_{o\!f\!f}$ moves toward the origin as the power increases and the SDE is largest near this node~\cite{souto2023tuning}. In device B, there is even no obvious SDE near zeroth step's first closing point, indicating that contribution from the RSJ mechanism is small. The difference between the experimental and RSJ results indicates a new origin of microwave-assisted SDE.


{\em Figure~\ref{fig_magnet} description.}---
We study the magnetic field response of device B in Fig.~\ref{fig_magnet}. The primary effect of increasing the magnetic field for a \red{Josephson junction} is a decrease in the critical current. This also increases the dimensionless frequency $h f/I_c R_n$, making Shapiro step oscillations closer to Bessel functions around constant currents [Figs.~\ref{fig_magnet}(a) and \ref{fig_magnet}(g)]. Here $h$ is Planck's constant, $f$ is the microwave frequency, $R_n$ is the normal state resistance. In Figs.~\ref{fig_magnet}(a)-\ref{fig_magnet}(c) where the magnetic field is negative, the zeroth-step oscillation nodes at high powers ($> 10$~dBm) fall on the right side of the origin. 
At zero magnetic field [Figs.~\ref{fig_magnet}(d)], these nodes are close to $I = 0$. In Figs.~\ref{fig_magnet}(e)-\ref{fig_magnet}(g) where the magnetic field is positive, the high-power nodes fall onto the left side of the origin, indicating the sign-reversal of $I_{o\!f\!f}$ when the direction of the magnetic field changes. Fig.~\ref{fig_magnet}(h) shows the evolution of Shapiro steps in the magnetic field at a fixed power. The zeroth step is inversion symmetric about the origin, shifting in the negative direction when $B$ is positive, and vice versa. $|I_{o\!f\!f}|$ first increases, reaching a maximum near $\pm 0.2$~T, then decreases, and finally vanishes near $\pm 0.8$ T which are close to critical fields.

\begin{figure}[H]\centering
\includegraphics{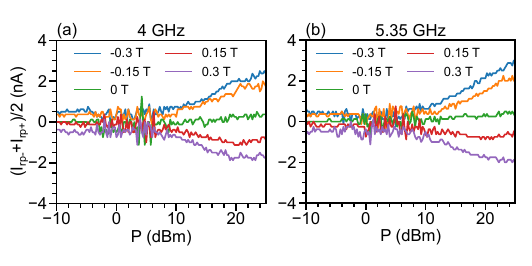}
\caption{\label{figS_analyse_retrapping} 
Supplementary data for Fig.~\ref{fig_analyse}. Here we calculate the offset current using the retrapping currents, $(I_{rt-} + I_{rt+})/2$.
}
\end{figure}

{\em Figure~\ref{fig_analyse} description.}---
The extracted offset current $I_{o\!f\!f}$ of device B is depicted in Fig.~\ref{fig_analyse}. $I_{o\!f\!f}$ is not symmetric between -0.15 T and 0.15 T, neither is a constant zero at 0~T. The asymmetry is consistent with the zero-field difference between $I_{rt-}$ and $I_{rt+}$ when the microwave is switched off (Fig.~\ref{figS_dev23_diode}). We attribute it to the zero-field SDE casued by trapped fluxes in the device. The SDE is enhanced in microwave irradiation. Here $I_{o\!f\!f}$ is defined as $(I_{sw-}+I_{sw+})/2$ and $(I_{rt-}+I_{rt+})/2$ gives similar values with less fluctuation (Fig.~\ref{figS_analyse_retrapping}). $I_{o\!f\!f}$ curves are more symmetric between -0.3 T and 0.3 T, and roughly increases linearly with the dBm power above 5 dBm, i.e., after the zeroth step oscillation's first node. 
The $I_{o\!f\!f}$ dependence on $P$ is opposite to the RSJ scenario where $I_{o\!f\!f}$ is finite without microwave and moves towards zero as the power increases~\cite{sm}.

{\em Discussion.}---
The dependence of $I_{o\!f\!f}$ on $P$ indicates different origins of the ideal SDE or USC observed in our work from the proposal in Ref.~\cite{souto2023tuning}. A possible mechanism is the nonequilibrium distribution of quasiparticle states, which is common in \red{Josephson junction}s at a finite voltage or in microwave irradiation~\cite{dubos2001coherent, dartiailh2021missing}. Nonequilibrium distribution causes the time-reversal symmetry breaking which is necessary for nonreciprocal behaviors~\cite{Linke_1998,PhysRevB.61.15914,sodemann2015quantum}. 
For bilayer superconducting systems, an in-plane offset current is predicted to be generated by the nonequilibrium steady state induced by an out-of-plane electric field~\cite{daido2023unidirectional}. 
\red{Refs.~\cite{Trahms2023nature, Steiner2023PRL} reported a new SDE mechanism by including asymmetric dissipative current and noise in the resistively and capacitively shunted junction model. Whether this effect is enhanced in the presence of an AC driving deserves further studies.}

\begin{figure}[H]\centering
\includegraphics[scale=0.95]{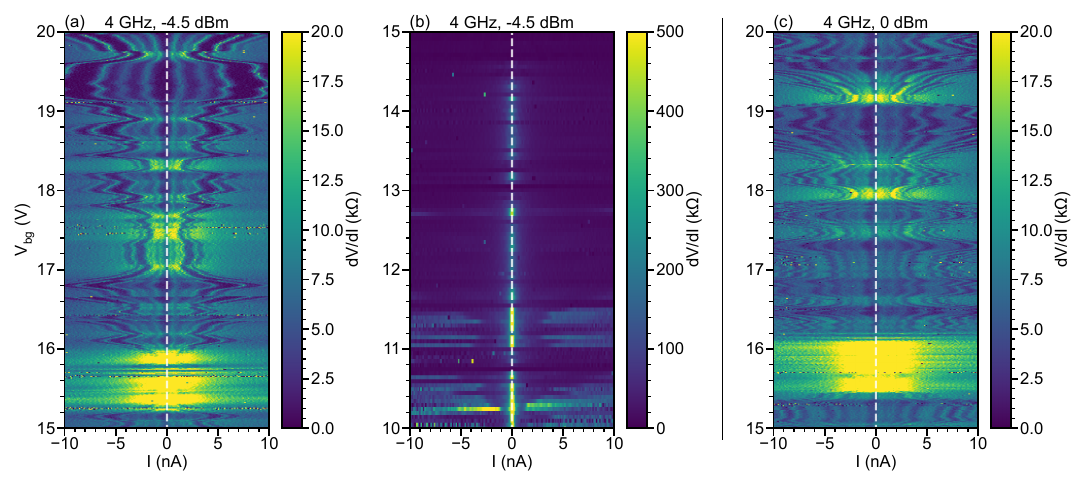}
\caption{\label{figS_chip24_gate} 
\red{Gate-voltage dependence of the zero-voltage step at fixed microwave power and magnetic field. The microwave frequency is 4 GHz.
(a) and (b) -4.5 dBm. (c) 0 dBm. The hysteresis behavior about the current-scan direction vanishes at both power values, so the asymmetrical zero voltage step indicates the SDE directly.
Positive and negative switching currents oscillate following similar patterns but with different oscillation amplitudes, indicating a gate-dependent offset current in the zero-voltage step. This is observed at -4.5 dBm (a) and is more obvious at 0 dBm where the positive switching current is close to zero (c).
At gate voltages below 15 V and microwave power -4.5 dBm, $dV/dI$ peaks are observed at $I = 0$ instead of zero-voltage steps (b). These peaks form a line along $I = 0$, indicating that there is no current offset due to leakage current or systematic errors.
The magnetic field is 0.15~T along the C-axis.}
}
\end{figure}

An alternative explanation is the rectifying effect due to non-linear components in series or parallel to the junction. Examples are Schottky barriers formed on metal-semiconductor interfaces and quantum dot states that are common in nanowire \red{Josephson junction}s~\cite{Fan_2016,C5NR04273A,chen2019ubiquitous,doi:10.1126/science.abf1513,levajac2023supercurrent}. This explanation may be consistent with the observation that $I_{o\!f\!f}$ increases with the microwave power. However, neither the series condition nor the parallel condition gives rise to the USC in the current bias scenario. For the series condition, the rectifying effect offsets the voltage instead of the current. For the parallel condition, measured $|I_{sw-}|$ and $|I_{sw+}|$ do not decrease, so switching currents can not be zero or reach the USC regime. We note that any non-zero $I_{o\!f\!f}$ can be regarded as a ``rectifying effect" phenomenologically, including the RSJ and nonequilibrium transport scenarios discussed in previous paragraphs. It is the origin of the nonlinearity that matters. The sign of $I_{o\!f\!f}$ is determined by whether the field and the DC current are parallel or anti-parallel in both devices, which prefers an intrinsic origin. \red{Leakage current or DC offset from the measurement setup is also unlikely since the offset in current vanishes when the zero-voltage step is suppressed by either the magnetic field or the gate voltage [Figs.~\ref{fig_magnet}(h) and \ref{figS_chip24_gate}]}.

In summary, the microwave-assisted USC observed in Al-InAs nanowire-Al junctions manifests similar signatures to the simulation based on the RSJ model, but can not be fully explained by the latter. While the origin of the microwave-assisted USC remains to be understood, nonequilibrium transport may play a role in this periodically driven system. 
Accidental zero-field asymmetry between retrapping currents also deserves future study in the context of zero-field SDE. The high-quality Josephson system used in this experiment provides a simple and tunable platform for studying the interplay between nonreciprocal and nonequilibrium physics. 

{\em Data availability.}---
Data and processing code are available at Ref.~\cite{zenodo}.

\begin{acknowledgments}
{\em Acknowledgements.}---
This work is supported by the NSFC (Grant Nos. 92165208, 11874071, 92365103, 12374480, 12374459, 61974138 and 92065106). D.P. acknowledges the support from Youth Innovation Promotion Association, Chinese Academy of Sciences (Nos. 2017156 and Y2021043).

\end{acknowledgments}

\bibliographystyle{apsrev4-2}
\bibliography{ref.bib}

\clearpage
\beginsupplement

\begin {center}
\textbf{\large{
Supplementary Materials: Microwave-assisted unidirectional superconductivity in Al-InAs nanowire-Al junctions under magnetic fields
}
}
\end {center}

\section{Device and measurement information}

\begin{table}[H]
\caption{\label{tab:dev_info} Device and cooldown reference. \red{Devices A and B are measured with a single axis magnet. Device C is measured with a three-axes magnet.}}
\begin{ruledtabular}
\begin{tabular}{lllll}
Name & Chip & Device & Nanowire batch & Cooldown \\
\hline
Device A & 20230111 Al-InAs-Chip-16 &  JJ-right-5 & InAs-epi-Al-2852  & Triton XL 2023-07-13 to 2023-09-13 \\
Device B & 20230711 Al-InAs-Chip-23 & JJ-left-8 & InAs-epi-Al-2852 & Triton XL 2023-07-13 to 2023-09-13\\
\red{Device C} & \red{20240131 Al-InAs-Chip-24} & \red{JJ-left-1} & \red{InAs-epi-Al-2852} & \red{Toploading 2024-03-30 to 2024-04-30}\\
\end{tabular}
\end{ruledtabular}
\end{table}

\begin{figure}[H]\centering
\includegraphics{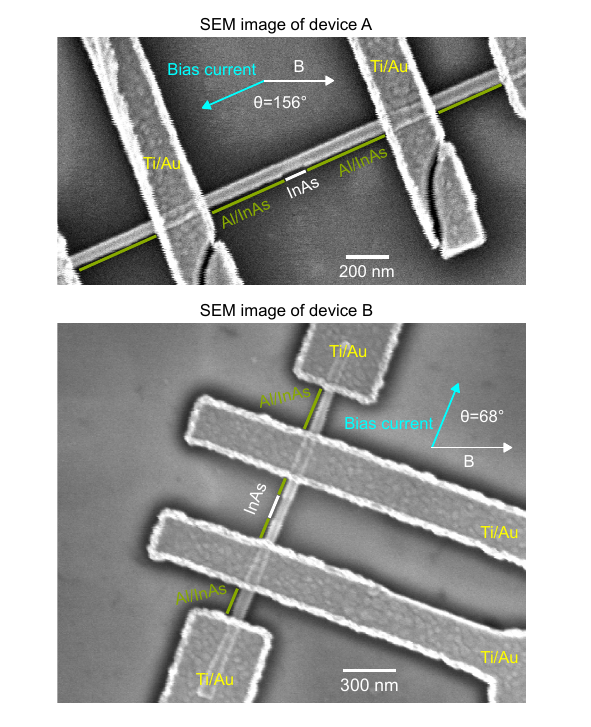}
\caption{\label{figS_SEM} 
SEM image of device A and device B. The section labeled ``Al/InAs” is the InAs nanowire covered with an epitaxial Al film. The section labeled ``InAs" is where the Josephson junction is. Ti/Au electrodes are 5/90 nm in thickness. The positive magnet field direction and positive bias current direction are indicated by white and blue arrows.
}
\end{figure}

\begin{figure}[H]\centering
\includegraphics{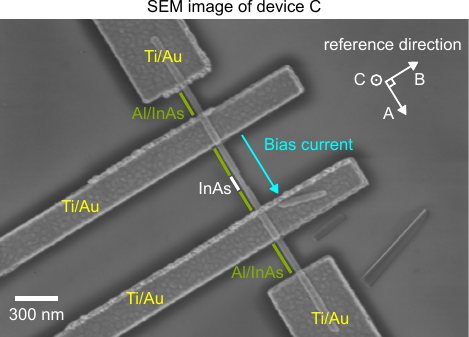}
\caption{\label{figS_chip24_L1_SEM} 
\red{SEM image of device C. Annotations are similar to those in Fig.~\ref{figS_SEM}. This device is measured in a dilution fridge equipped with a three-axis magnet.}
}
\end{figure}

\begin{figure}[H]\centering
\includegraphics[scale=0.7]{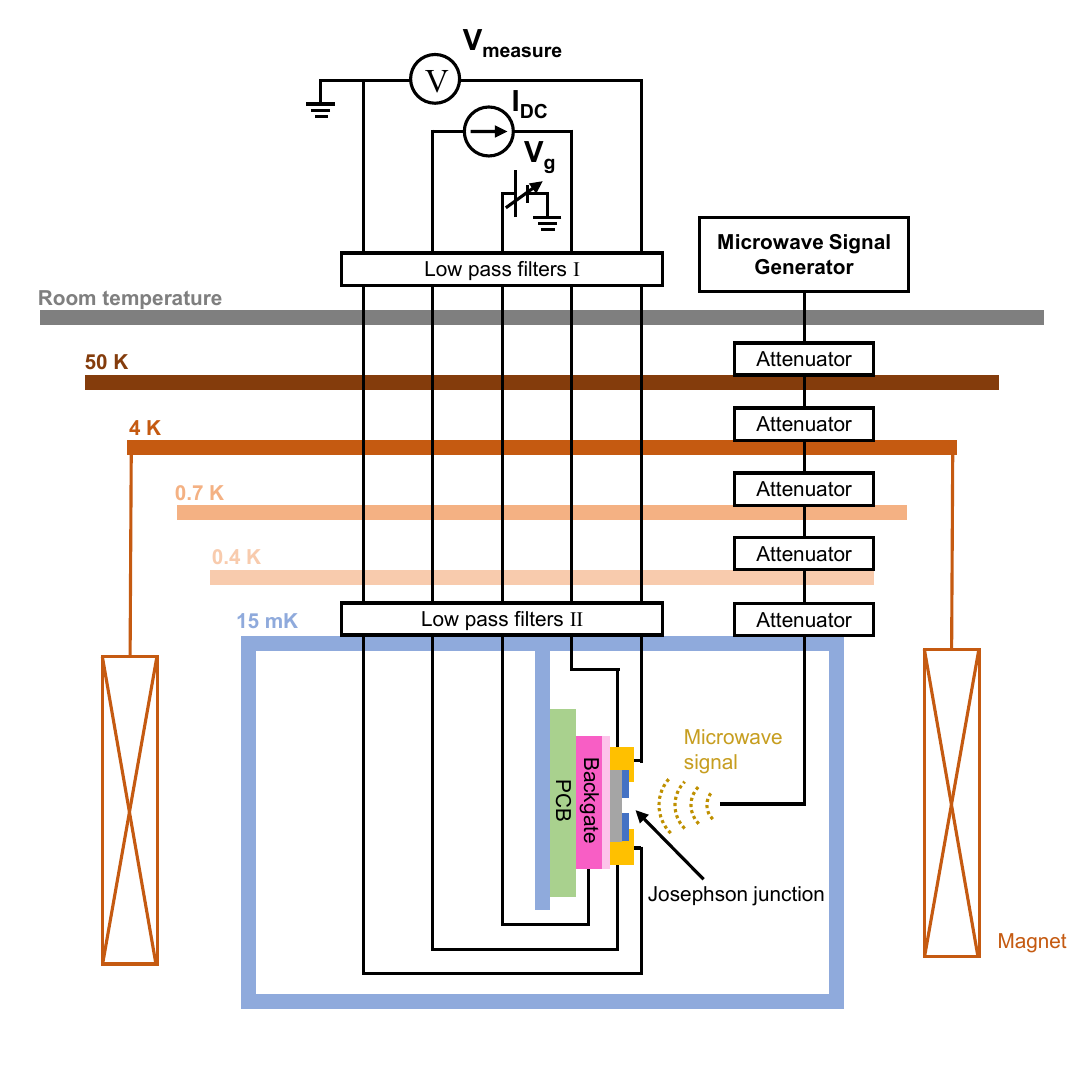}
\caption{\label{figS_Circuit_element_diagram_XL} 
\red{Measurement setup for devices A and B. The total attenuation in the microwave line is 29 dB. For device C, the measurement takes place in a different dilution refrigerator with a similar setup, except that the magnet has three axes instead of a single axis, the device is mounted vertically in a downward direction, and the total attenuation in the microwave line is 23~dB.}
}
\end{figure}


\section{RSJ simulation}

\section{Extended data from devices A and B}

\begin{figure}[H]\centering
\includegraphics{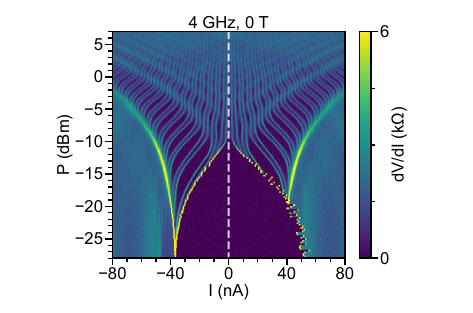}
\caption{\label{figS_dev16_4g_0T} 
Supplementary data to Fig.~\ref{fig_overview}(c). Lines in Fig.~\ref{fig_overview}(c) are extracted from this dataset.
}
\end{figure}

\begin{figure}[H]\centering
\includegraphics{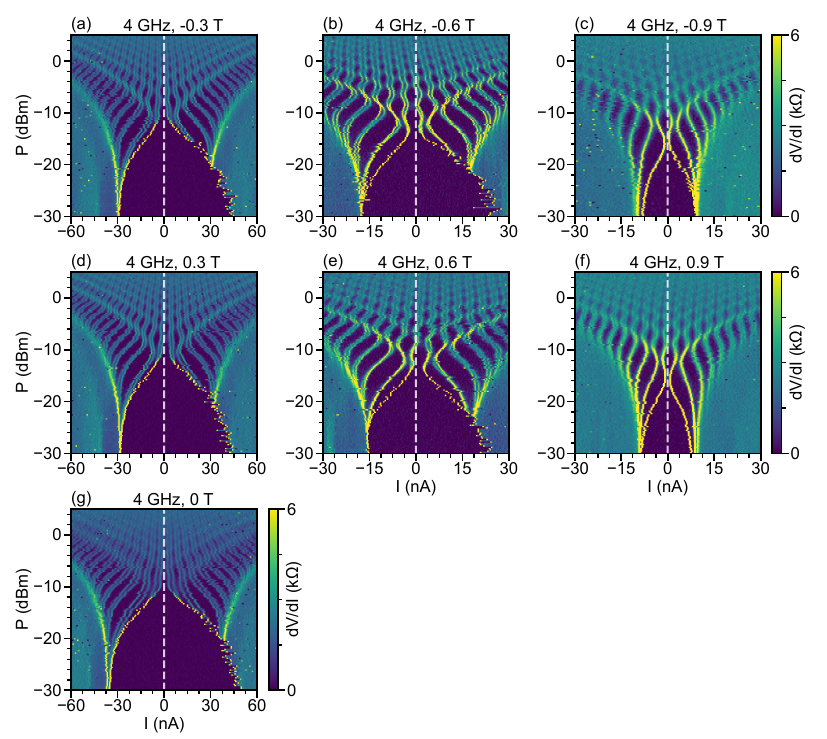}
\caption{\label{figS_dev16_4g_magnet} 
Additional data for device A in a variety of magnetic fields showing the  microwave-assisted SDE. $f = 4$ GHz. The back gate voltage is -0.6 V. The current is scanned in the positive direction in all panels.
}
\end{figure}




\begin{figure}[H]\centering
\includegraphics{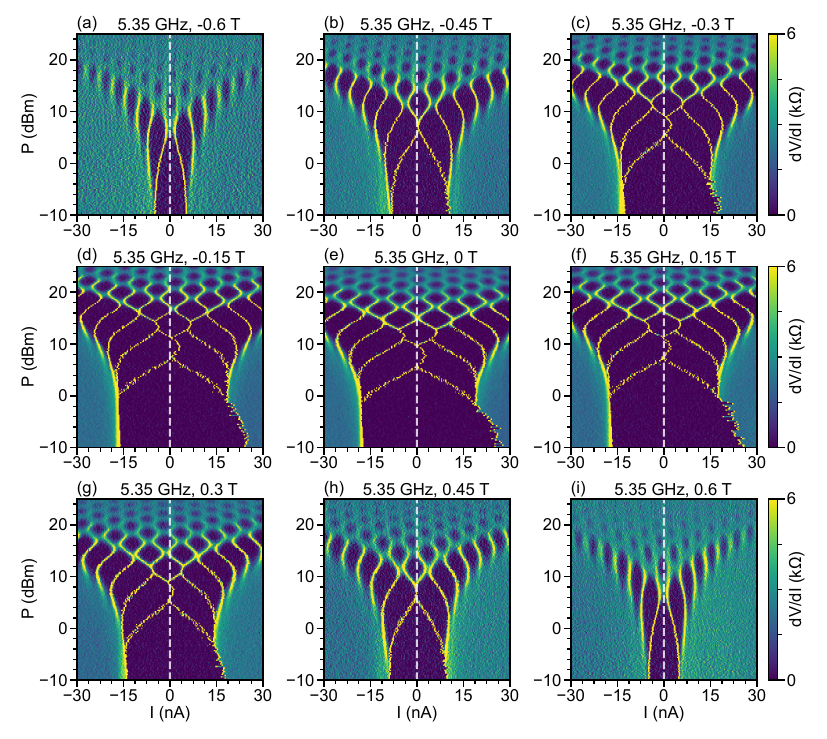}
\caption{\label{figS_dev23_5.35g_magnet} 
Additional data for device B in a variety of magnetic fields showing the microwave-assisted SDE. $f = 5.35$ GHz. The back gate voltage is 27 V. The current is scanned in the positive direction in all panels.
}
\end{figure}


\begin{figure}[H]\centering
\includegraphics{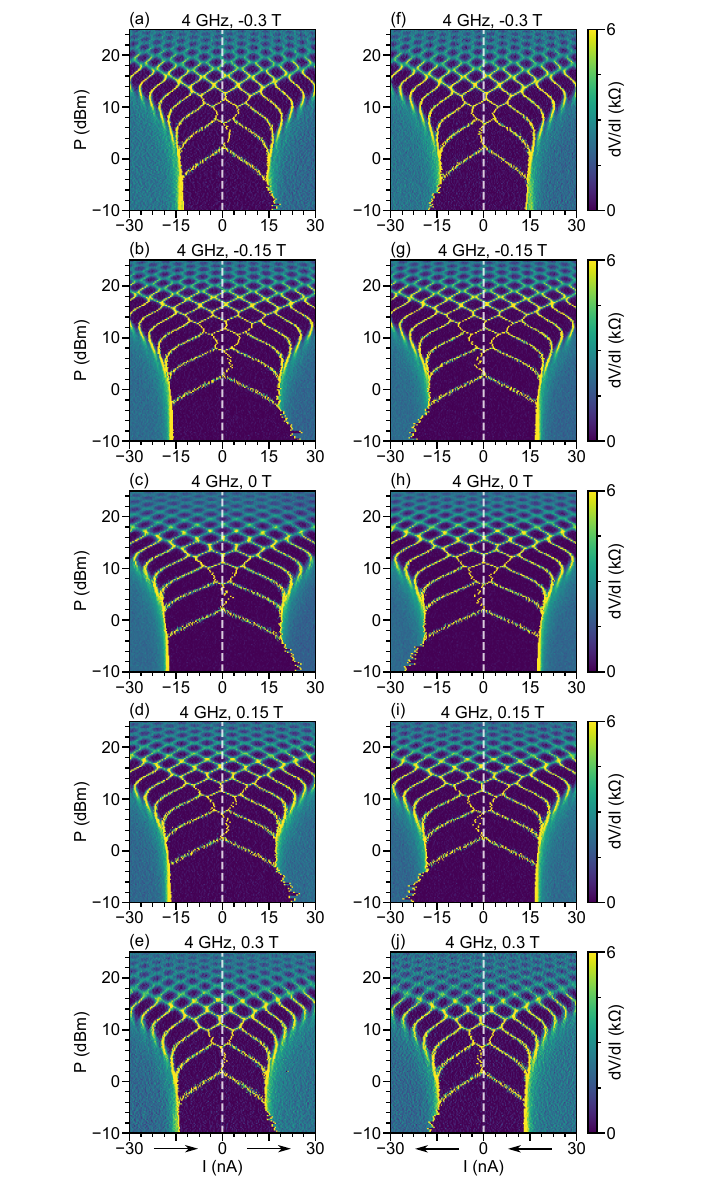}
\caption{\label{figS_dev23_4g_direction} 
Current scan direction dependence of the microwave-assisted SDE in device B at 4 GHz. (a-j) Differential resistance $dV/dI$ as a function of the microwave power $P$ and the bias current $I$. The microwave frequency and the magnetic field are indicated at the top of each panel. The back gate voltage is 27 V. Horizontal arrows indicate the scan direction of $I$.
}
\end{figure}


\begin{figure}[H]\centering
\includegraphics{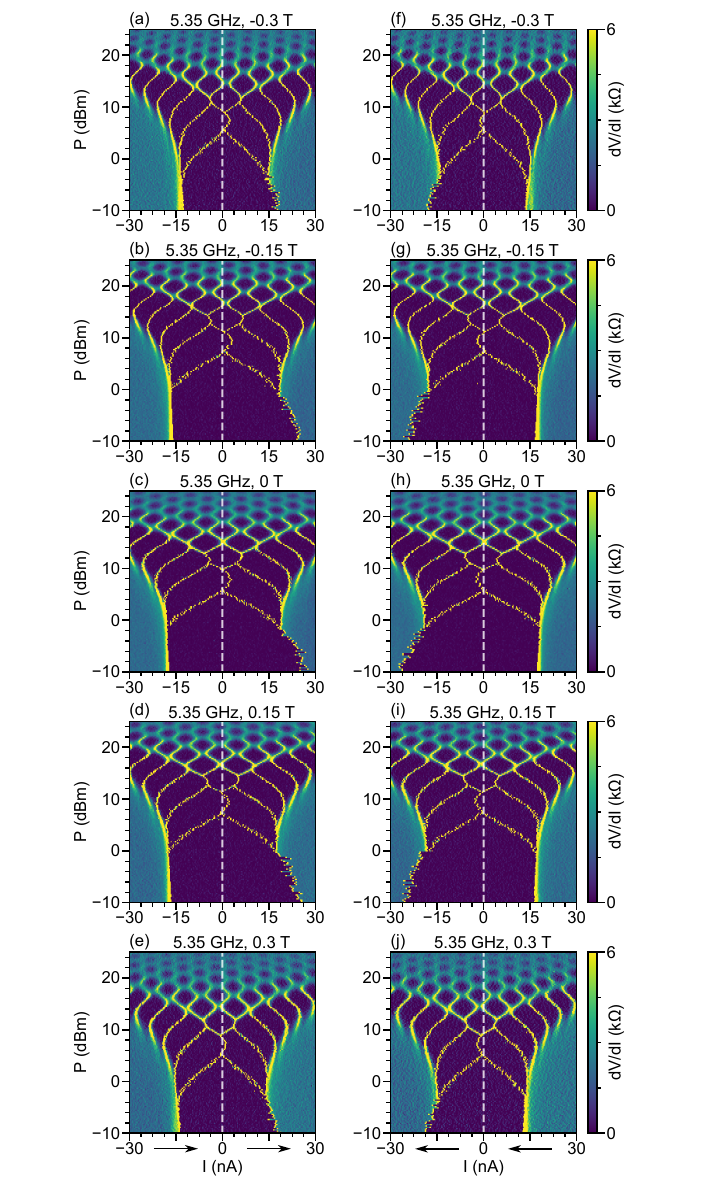}
\caption{\label{figS_dev23_5.35g_direction} 
Current scan direction dependence of the microwave-assisted SDE in device B at 5.35 GHz. (a-j) Differential resistance $dV/dI$ as a function of the microwave power $P$ and the bias current $I$. The microwave frequency and the magnetic field are indicated at the top of each panel. The back gate voltage is 27 V. Horizontal arrows indicate the scan direction of $I$.
}
\end{figure}

\begin{figure}[H]\centering
\includegraphics{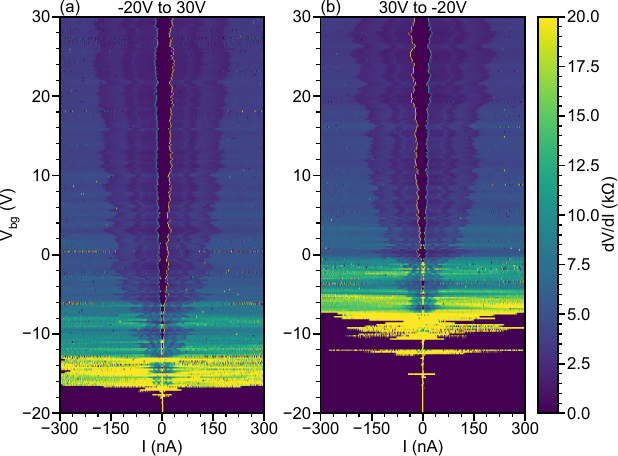}
\caption{\label{figS_chip23_Vbg_I} 
\red{Gate-voltage dependence of Device B. The microwave is off. The magnetic field is zero. (a) $V_{bg}$ is scanned from -20~V to 30~V. $I$ is scanned from -300~nA to 300~nA. (b) $V_{bg}$ is scanned from 30~V to -20~V. $I$ is scanned from 300~nA to -300~nA. We note that due to the hysteresis about the gate voltage scan direction, the result is history dependent. The switching current and the normal state resistance at $V_{bg} = 27$~V in panel (a) are 25 nA and 3.5~k$\Omega$, respectively.}
}
\end{figure}

\red{\section{Extended data from device C}}

\begin{figure}[H]\centering
\includegraphics[scale=0.95]{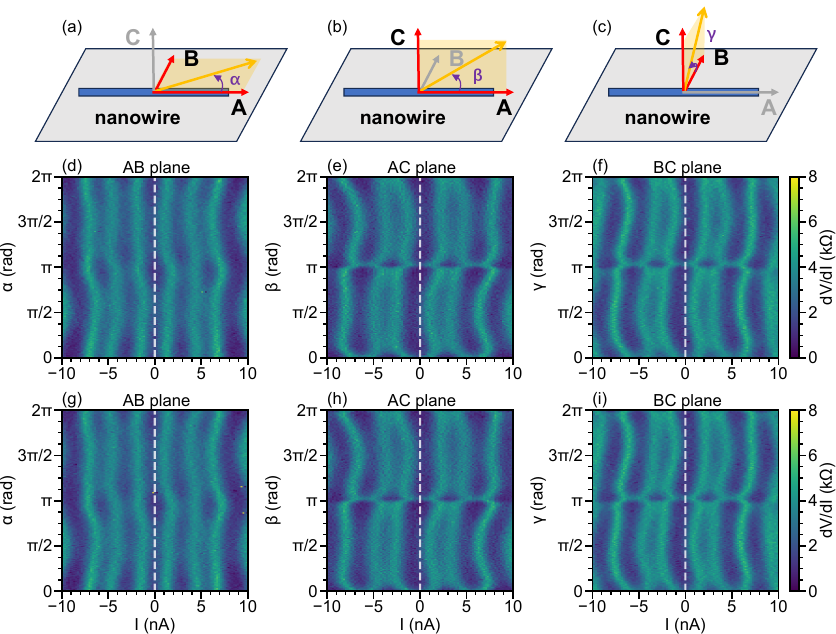}
\caption{\label{figS_chip24_B_rotation}
\red{Magnetic field orientation dependence of the microwave-assisted SDE at a finite microwave power (device C).
(a)-(c) Schematic diagrams of the field rotation. The yellow arrow represents the magnetic field which is rotated in the AB, AC, and BC planes, respectively. The magnitude of the magnetic field is fixed at 0.15 T.
(d)-(f) $dV/dI$ as a function of rotation angle and DC current. The offset of the zero-voltage step varies as a function of the angle. The amplitude of the variation is larger when the field is rotated in the AC and BC planes and smaller when the field is rotated in the AB plane. Assuming the offset originates from SDE due to spin-orbit coupling, the observations indicate a spin-orbit field perpendicular to the substrate.
Such a spin-orbit field is possible if the built-in electric field, determined by the orientation of the aluminum shell on the nanowire, is along the B-axis~\cite{mazur2022gate, zhang2022evidence}.
The current is scanned in the positive direction. 
(g)-(i) Similar to (d)-(f) except that the current is scanned in the negative direction. No obvious hysteresis is observed compared to (d)-(f).
The microwave frequency and power are 4~GHz and 0~dBm, respectively. The backgate voltage is 20 V. For field rotation without microwave irradiation, see Fig.~\ref{figS_chip24_B_rotation_diode}. For more power values, see Fig.~\ref{figS_chip24_B_rotation_six2d}.} 
\redd{Here it should be noted that power values from experiments on different devices, fridges, or microwave frequencies cannot be directly compared. This is because the RF signal is attenuated differently due to the fact that attenuators, circuit setup details, as well as the gap between the antenna and the Josephson junction (Fig.~\ref{figS_Circuit_element_diagram_XL}) in different fridges and different experiments can be very different. The actual attenuation depends on many factors and varies from experiment to experiment. A practical reference for comparison is the power value where the lobe of the zeroth Shapiro step closes as the microwave power increases. See Figs.~\ref{fig_freq_and_dev} and \ref{figS_chip24_B_rotation_six2d} for typical values of this reference point in devices A, B, and C.}
}
\end{figure}

\begin{figure}[H]\centering
\includegraphics[scale=0.95]{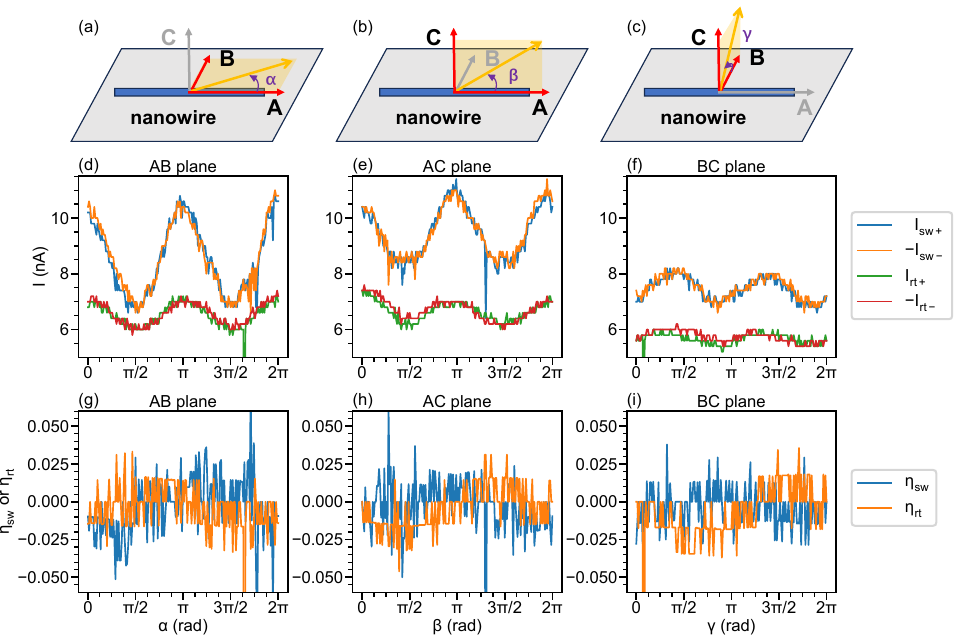}
\caption{\label{figS_chip24_B_rotation_diode} 
\red{Magnetic field rotation in device C without microwave irradiation. 
(a)-(c) Schematic diagrams of the field rotation experiment. The yellow arrow represents the magnetic field which is rotated in the AB, AC, and BC planes, respectively. The magnitude of the magnetic field is fixed at 0.15 T.
(d)-(f) Extracted switching and retrapping currents.
(g)-(i) Calculated diode efficiencies from switching [$\eta_{sw} = (|I_{sw+}|-|I_{sw-}|)/(|I_{sw+}|+|I_{sw-}|)$] and retrapping [$\eta_{rt} = (|I_{rt+}|-|I_{rt-}|)/(|I_{rt+}|+|I_{rt-}|)$] currents. $\eta_{rt}$ manifests a weak variation as the angle changes, reaching its maximum (minimum) when the field is anti-parallel (parallel) to the C-axis. This is consistent with the angle dependence under microwave irradiation (Fig.~\ref{figS_chip24_B_rotation}).
The backgate voltage is 20 V.}
}
\end{figure}


\begin{figure}[H]\centering
\includegraphics[scale=0.95]{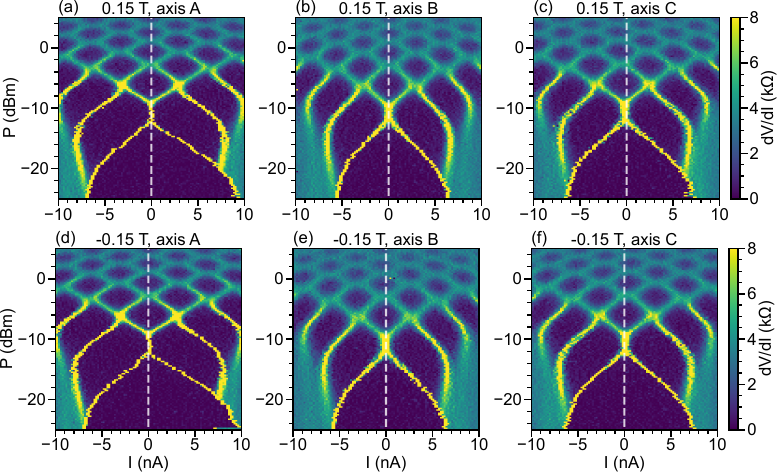}
\caption{\label{figS_chip24_B_rotation_six2d} 
\red{Microwave-assisted SDE at typical magnetic field directions. (a)-(c) Field directions are parallel to axes A, B, and C, respectively.
(d)-(f) Field directions are anti-parallel to axes A, B, and C, respectively.
Consistent with previous results, the microwave-assisted SDE is weak (strong) when the field is along the A-axis or B-axis (C-axis). The backgate voltage is 20 V.}
}
\end{figure}

\begin{figure}[H]\centering
\includegraphics[scale=0.95]{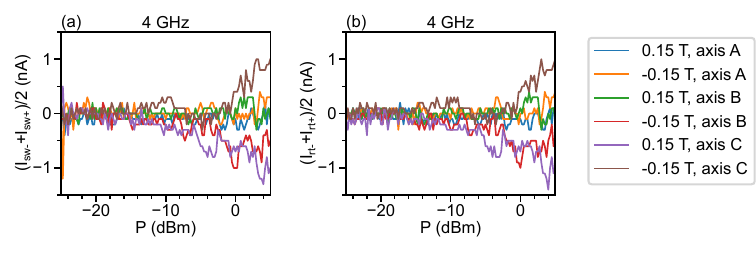}
\caption{\label{figS_DeviceC_rotation_analyse_six} 
\redd{Extracted offset currents at typical magnetic field directions. (a) $(I_{sw-} + I_{sw+})/2$, (b) $(I_{rt-} + I_{rt+})/2$. Switching and retrapping currents are extracted from datasets including those in Fig.~\ref{figS_chip24_B_rotation_six2d} and in both current-scan directions. The backgate voltage is 20 V.}
}
\end{figure}

\begin{figure}[H]\centering
\includegraphics[scale=0.95]{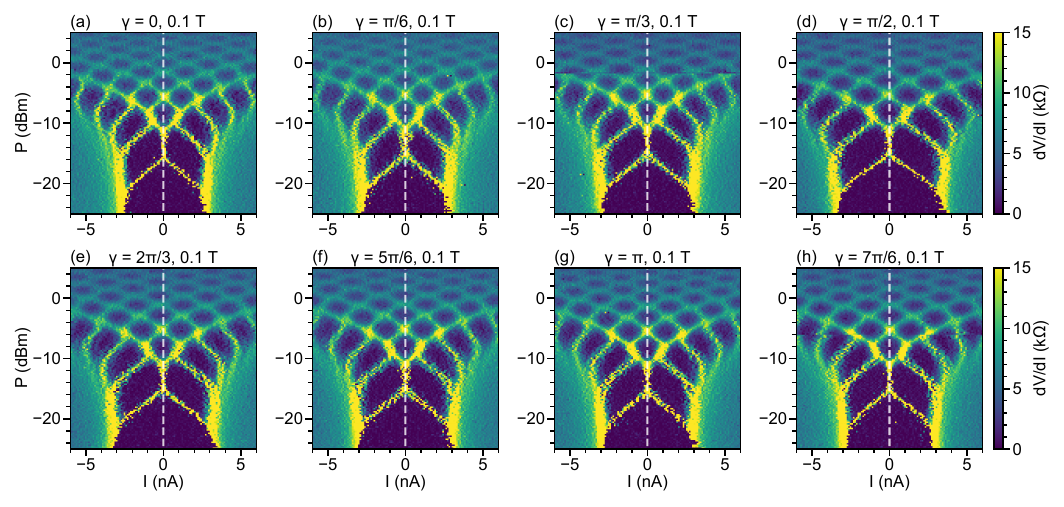}
\caption{\label{figS_DeviceC_rotation_210_2D} 
\redd{Field rotation in the BC plane. $\gamma$ is the angle between the magnetic field and B-axis (Fig.~\ref{figS_chip24_B_rotation_diode}(c)). The difference in critical currents at -25 dBm compared to those in Fig.~\ref{figS_chip24_B_rotation_six2d} is due to the device change after thermal cycles and gate voltage cycles. The backgate voltage is 20 V.}
}
\end{figure}

\begin{figure}[H]\centering
\includegraphics[scale=0.95]{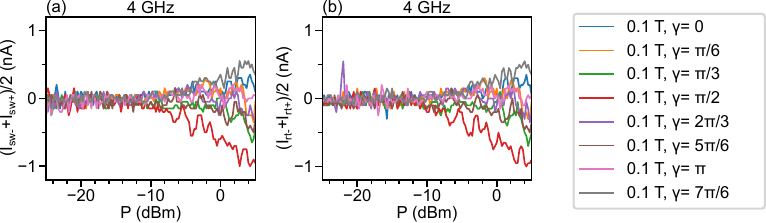}
\caption{\label{figS_DeviceC_rotation_analyse_210} 
\redd{Extracted offset currents for magnetic field directions rotated in the BC plane. (a) $(I_{sw-} + I_{sw+})/2$, (b) $(I_{rt-} + I_{rt+})/2$. Switching and retrapping currents are extracted from datasets including those in Fig.~\ref{figS_DeviceC_rotation_210_2D} and in both current-scan directions. For field rotation experiments at a fixed power, see Fig.~\ref{figS_chip24_B_rotation}.}
}
\end{figure}



\begin{figure}[H]\centering
\includegraphics{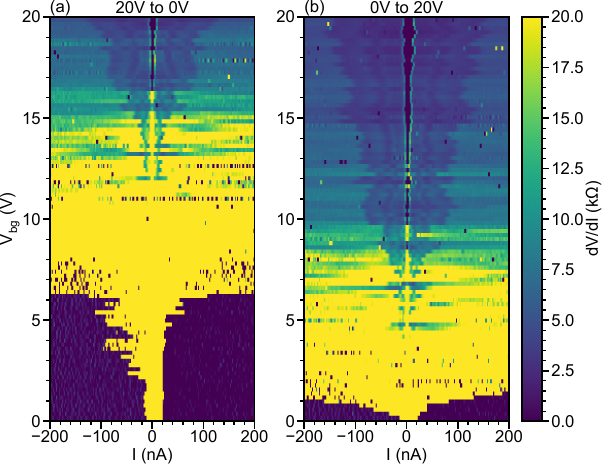}
\caption{\label{figS_chip24_Vbg_I} 
\red{Gate-voltage dependence of Device C with microwave off. (a) $V_{bg}$ is scanned from 20~V to 0~V. (b) $V_{bg}$ is scanned from 0~V to 20~V. The magnetic field is zero. $I$ is scanned from -200~nA to 200~nA. We note that due to the hysteresis about the gate voltage scan direction, the result is history dependent. The switching current and the normal state resistance at $V_{bg} = 20$~V in panel (b) are 12~nA and 4.5~k$\Omega$, respectively.}
}
\end{figure}

\begin{figure}[H]\centering
\includegraphics{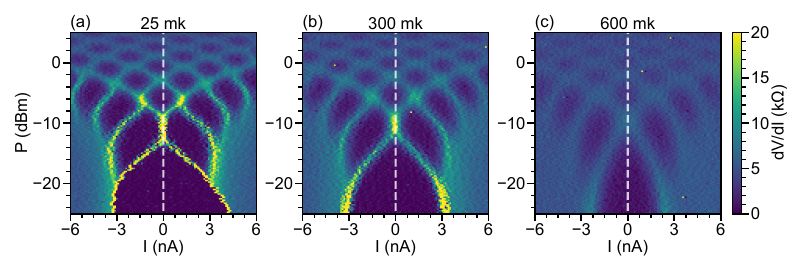}
\caption{\label{figS_chip24_Temp} 
\red{Temperature dependence of Device C. The microwave frequency is 4~GHz. The magnetic field is 0.15~T along the C-axis. The temperature is noted at the top of each panel. The microwave-assisted SDE persists at 300 mK while the hysteresis behavior at low powers is strongly suppressed.}
}
\end{figure}

\end{document}